\documentclass[
prd
,showpacs,amssymb,superscriptaddress,aps,
nofootinbib
]{revtex4-2}
\usepackage{graphicx}
\usepackage{color}
\input{epsf}

\usepackage{amsmath,amssymb}
\usepackage{bm}
\usepackage{times}
\usepackage{ulem}

\newcommand{\dalm}{\kern1pt\vbox{\hrule height 0.9pt\hbox{\vrule width 0.9pt
\hskip 2.5pt\vbox{\vskip 5.5pt}\hskip 3pt\vrule width 0.3pt}\hrule height 0.3pt}
\kern1pt}

\begin{document}



\title{The $g$-mode frequencies in cold neutron stars and nuclear saturation parameters}

\author{Hajime Sotani}
\email{sotani@yukawa.kyoto-u.ac.jp}
\affiliation{Department of Mathematics and Physics, Kochi University, Kochi, 780-8520, Japan}
\affiliation{RIKEN Center for Interdisciplinary Theoretical and Mathematical Sciences (iTHEMS), RIKEN, Wako 351-0198, Japan}
\affiliation{Kochi University of Technology, Kochi, 780-8515, Japan.}
\affiliation{Theoretical Astrophysics, IAAT, University of T\"{u}bingen, 72076 T\"{u}bingen, Germany}

\author{Hajime Togashi}
\affiliation{Department of Physics, Kyoto University, Kyoto 606-8502, Japan}


\date{\today}

\begin{abstract}
Oscillation frequencies excited in neutron stars are crucial for extracting their interior properties. In addition to the fundamental and pressure modes, the gravity ($g$-) modes can be excited even in zero-temperature stellar models due to the composition gradient. In this study, we systematically study the $g$-mode frequencies, focusing on the nucleonic equation of state. Then, we can derive an empirical relation for the 1st $g$-mode frequencies as a function of the stellar compactness and the combination of the nuclear saturation parameters, $\eta_0 \equiv L/K_0$, where $K_0$ and $L$ denote the incompressibility of symmetric nuclear matter and the density dependence of the nuclear symmetry energy, respectively. If an observed 1st $g$-mode frequency significantly deviates from our empirical relation, it may indicate the emergence of additional degrees of freedom or new compositions inside the star.
\end{abstract}

%
\maketitle


\section{Introduction}
\label{sec:I}

Neutron stars, which are produced via a supernova explosion at the last moment of a massive star's life, achieve extreme conditions that are virtually impossible to reproduce on Earth~\cite{ST83}. For example, the density inside the star easily exceeds the nuclear saturation density and may become several times larger than the saturation density, depending on the stiffness of the equation of state (EOS) for neutron star matter. The gravitational and magnetic fields also become much stronger than those observed in our solar system. Thus, it can be expected that one can probe physics under such extreme states through the observations of neutron stars and their phenomena. In particular, the knowledge of the EOS for higher-density regions is still poor, even though many EOS have been proposed theoretically. This is due to the difficulty in obtaining information on nuclear characteristics in high-density regions through ground-based experiments, stemming from nuclear saturation properties. Constraints on the EOS represent one of the most important unsolved issues in neutron star physics.

Observational constraints on neutron star mass and radius provide crucial information for selecting the EOS, especially for high-density regions. In practice, the discoveries of massive neutron stars have excluded the soft EOS, with which the expected maximum mass does not reach the observed mass~\cite{D10,A13,F21}. Meanwhile, the careful observations of the pulsar light curve principally tell us the stellar compactness, i.e., the ratio of mass to radius, e.g.,~\cite{PDC18,Leahy03,PG03,PO14,SM18,S20}. This is because the paths of photons emitted from the surface of neutron stars are bent by the gravitational fields induced by the neutron star due to a relativistic effect. Actually, the neutron star mass and radius for PSR J0030+0451~\cite{Riley19,Miller19,Blaschke20} and for PSR J0740+6620~\cite{Riley21,Miller21,Dittmann24} are constrained by the X-ray observations with the Neutron Star Interior Composition Explorer (NICER) on the International Space Station. The gravitational waves also become a new tool to obtain astronomical information. The GW170817~\cite{GW170817} gives us a constraint on the neutron star tidal deformability, which leads to the radius constraint, i.e., the radius for a $1.4M_\odot$ neutron star should be less than 13.6 km~\cite{Annala18}. Of course, ground-based experiments are crucial for restricting the EOS in the low-density region~\cite{SNN22,SO22,SN23}.

As well as the neutron star mass and radius, the frequencies detected from a neutron star are another crucial source of information for obtaining the stellar properties. Since the object has eigenfrequencies that depend on its interior properties, one could extract stellar information by identifying the observed frequencies with specific modes as an inverse problem, provided one had previously derived the relation between frequencies and stellar properties. This technique is known as asteroseismology, which is in the same way as seismology on Earth and helioseismology on the Sun. In practice, by identifying the magnetar quasi-periodic oscillations (QPOs) with the crustal torsional oscillations, the crust EOS and neutron star mass and radius are constrained, e.g.,~\cite{GNHL2011,SNIO2012,SIO2016,SKS23,Sotani24a}. In the same way, recently the mass and radius of a fast radio burst (FRB 20240114A) have also been estimated~\cite{SWC26}. Furthermore, once the gravitational waves from a neutron star are detected in the future, they also tell us the neutron star mass, radius, EOS, and/or rotational properties with the so-called gravitational wave asteroseismology, e.g.,~\cite{AK1996,AK1998,STM2001,SH2003,TL2005,SYMT2011,PA2012,DGKK2013,Sotani20b,Sotani21,SK21,Sotani23}. This also applies to supernova gravitational waves, e.g.,~\cite{FMP2003,FKAO2015,ST2016,ST2020a,SKTK2017,MRBV2018,SKTK2019,TCPOF19,SS2019,ST2020,STT2021,SMT24}.

The oscillation frequency excited in the neutron star corresponds to each physical process under consideration. Among a variety of modes, the fundamental ($f$-) and pressure ($p$-) modes, whose restoring force is the pressure gradient, have been well studied up to now. On the other hand, the studies of the gravity ($g$-) modes are relatively limited. Since the $g$-modes are excited due to the temperature and/or composition gradients, one can definitely see such modes in cooling or accreting neutron stars, e.g., \cite{KHA15,PAH16,SD22}. In addition, since the composition inside the star is generally not constant even for a cold neutron star, the $g$-modes can be excited due to the composition gradient \cite{Counsell25}. In such a situation, the sound velocity, $c_{\rm s}$, defined by the derivative of the pressure with respect to the energy density by fixing the composition fraction [Eq.~(\ref{eq:cs})] deviates from the velocity, $c_{\rm eq}$, given by the derivative of the pressure with respect to the energy density for $\beta$-equilibrium matter [Eq.~(\ref{eq:cs_eq})]. As a result, this difference becomes a source for the excitation of the $g$-modes, if the timescale of the weak interaction is longer than that of the stellar oscillations, \cite{Jaikumar21,Constantinou21,Zhao22,Tran23}. This deviation between two velocities, $c_{\rm s}$ and $c_{\rm eq}$, depends on the EOS; the resultant $g$-mode frequencies also depend on the EOS. However, up to now, a systematic study of how the $g$-mode frequencies are expressed by the EOS parameters has never been conducted. In this study, we focus on the nucleonic EOS and systematically study the dependence of the $g$-mode frequencies on the EOS.

This manuscript is organized as follows. In Sec.~\ref{sec:EOS}, we show the neutron star models with several EOS considered in this study and also briefly describe the definition of two velocities, $c_{\rm s}$ and $c_{\rm eq}$. In Sec.~\ref{sec:Oscillations}, we discuss the behavior of $g$-mode frequencies and their dependence on the EOS parameters. Then, in Sec.~\ref{sec:Dep_cs}, we discuss how the $g$-mode frequencies depend on the absolute value of the sound velocity. Finally, we conclude this study in Sec. \ref{sec:Conclusion}. Unless otherwise mentioned, we adopt geometric units in the following, $c=G=1$, where $c$ denotes the speed of light, and the metric signature is $(-,+,+,+)$.

\section{EOS and Neutron star models}
\label{sec:EOS}

In this study, we simply focus on static, spherically symmetric neutron star models as a background model. To describe such an object, the metric is given by 
\begin{equation}
  ds^2 = -e^{2\Phi}dt^2 + e^{2\Lambda}dr^2 + r^2 (d\theta^2 + \sin^2\theta d\phi^2), \label{eq:metric}
\end{equation}
where the metric functions, $\Phi$ and $\Lambda$, depend only on the radial coordinate, $r$, and $\Lambda$ is directly associated with the mass function, $m(r)$, which is the gravitational mass inside radius $r$, through 
\begin{equation}
  e^{-2\Lambda} = 1-\frac{2m}{r}. \label{eq:eLam}
\end{equation} 
The neutron star models can be constructed by solving the Tolman-Oppenheimer-Volkoff (TOV) equations with an appropriate EOS. In this study, we particularly adopt several EOS models with the Skyrme-type effective interaction, i.e.,  KDE0v~\cite{KDE0v}, SLy4~\cite{SLy4},  SkMp~\cite{SkMp}, SKa~\cite{SKa}, and SkI3~\cite{SkI3}; those based on the relativistic mean field approximation, i.e., G3 and IOPB-I~\cite{G3,IOPB,Parmar22}; and that derived using the variational method, i.e., Togashi~\cite{Togashi}. For the uniform core region, we calculate the EOSs for charge-neutral n, p, e, $\mu$ matter away from $\beta$ equilibrium using the corresponding nuclear interactions and construct tables suitable for evaluating both the equilibrium and frozen-composition sound speeds discussed below. For the non-uniform crust region, we adopt $\beta$-equilibrium crust EOSs~\cite{Parmar22,Togashi,GR15} that are consistent with each core EOS and connect them to the corresponding uniform-matter EOS.

\begin{figure}[tbp]
\begin{center}
\includegraphics[scale=0.6]{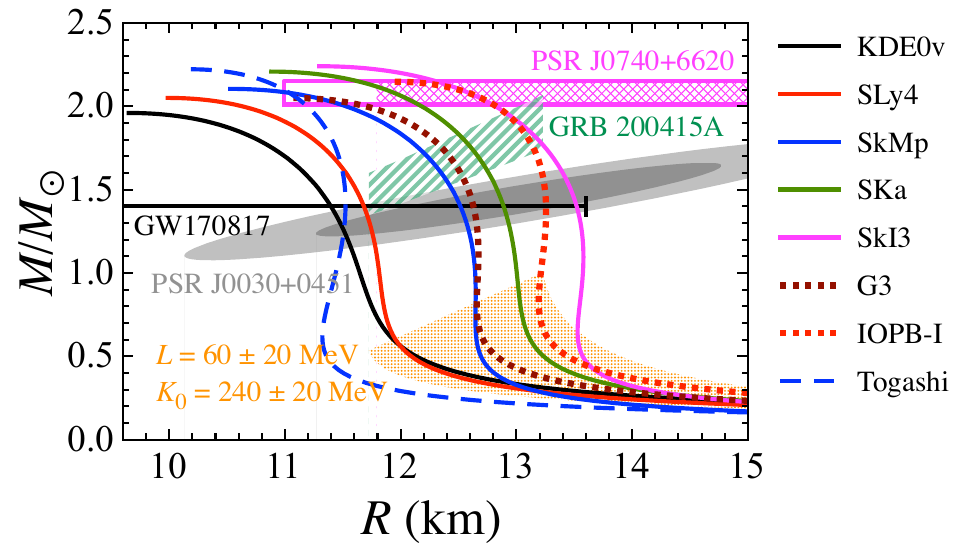} 
\end{center}
\caption{
Mass-radius relations for neutron star models constructed with several EOSs, where the solid, dotted, and dashed lines correspond to the EOSs with Skyrme-type effective interactions, those based on the relativistic mean field approximation, and the Togashi EOS, respectively. For reference, the constraints obtained by NICER observations for PSR J0030+0451 \cite{Riley19,Miller19,Blaschke20} and PSR J0740+6620 \cite{Riley21,Miller21,Dittmann24}, from the GW170817 event \cite{Annala18}, and with the high-frequency QPOs from GRB 200415A \cite{SKS23}, are also shown. In addition to the constraints from the astronomical observations, the neutron star mass and radius estimated with the mass formula for a low-mass neutron star \cite{SIOO14} are shown in the right-bottom region, assuming the fiducial values of $K_0$ and $L$, such as $K_0=240\pm 20$ MeV~\cite{Shlomo06} and $L=60\pm 20$~\cite{Li19}.  }
\label{fig:MR}
\end{figure}

The corresponding EOS parameters (the incompressibility $K_0$ and the density dependence of the nuclear symmetry energy $L$) are listed in Table~\ref{tab:EOS}. In this table, we also list the values of $\eta_0\equiv L/K_0$ described below; $\eta\equiv(K_0L^2)^{1/3}$ \cite{SIOO14}, which is a suitable parameter to characterize the low-mass neutron stars; and the maximum mass of the neutron star constructed with each EOS. The mass and radius of the neutron stars constructed with these EOSs are shown in Fig.~\ref{fig:MR}, where the solid lines denote the results with the EOSs with Skyrme-type interactions, the dotted lines denote the results with the EOSs based on the relativistic mean field approximation, and the dashed line denotes the result with the Togashi EOS.

\begin{table}
\caption{EOSs adopted in this study and corresponding nuclear parameters, $K_0$ and $L$, are the incompressibility and density dependence of the nuclear symmetry energy at the saturation density for symmetric nuclear matter, while $\eta_0\equiv L/K_0$ and $\eta\equiv(K_0L^2)^{1/3}$. In addition to the nuclear parameters, the maximum mass of the neutron star constructed with each EOS is listed in the last column.} 
\label{tab:EOS}
\begin {center}
\begin{tabular}{cccccc}
\hline\hline
EOS & $K_0$ (MeV) & $L$ (MeV) & $\eta_0$ & $\eta$ (MeV) & $M_{\rm max}/M_\odot$    \\
\hline
KDE0v   & 229 & 45.2 & 0.198 & 77.6 & 1.96  \\
SLy4    & 230 & 45.9 & 0.200 & 78.7 & 2.05  \\
SkMp    & 231 & 70.3 & 0.304 & 105  & 2.10  \\
SKa     & 263 & 74.6 & 0.283 & 114  & 2.21  \\
SkI3    & 258 & 101  & 0.389 & 138  & 2.24  \\
G3      & 244 & 49.3 & 0.202 & 84.0 & 2.05  \\   
IOPB-I  & 223 & 63.6 & 0.286 & 96.6 & 2.15  \\
Togashi & 245 & 34.4 & 0.141 & 66.1 & 2.22  \\
\hline \hline
\end{tabular}
\end {center}
\end{table}

In the same figure, we also show the constraints obtained from the astronomical observations:
the mass of the neutron star, PSR J0740+6620, is estimated to be $2.08^{+0.07}_{-0.07}M_\odot$ with 68.3\% credibility~\cite{A13,F21}; 
the mass and radius of the neutron stars, PSR J0740+6620 and PSR J0030+0451, are constrained by the NICER observations~\cite{Riley19,Miller19,Riley21,Miller21};
the $1.4M_\odot$ neutron star radius should be less than $13.6$ km from the gravitational wave observations in the GW170817 event~\cite{Annala18}; 
and the mass and radius constraints to explain the high frequencies of quasi-periodic oscillations observed from the magnetar GRB 200415A with the crustal torsional oscillations~\cite{SKS23}. 
In addition to the constraints obtained through the astronomical observations, the experimental constraints on the nuclear saturation parameters also give us the estimation of the neutron star mass and radius with a lower central density, which is located on the right-bottom region in Fig.~\ref{fig:MR}, using the low-mass neutron star mass formula proposed in Ref.~\cite{SIOO14} and assuming the fiducial values of $K_0=240\pm 20$ MeV~\cite{Shlomo06} and $L=60\pm 20$ MeV~\cite{Li19}.

\subsection{Sound velocity}
\label{sec:cs}

\begin{figure}[tbp]
\begin{center}
\includegraphics[scale=0.6]{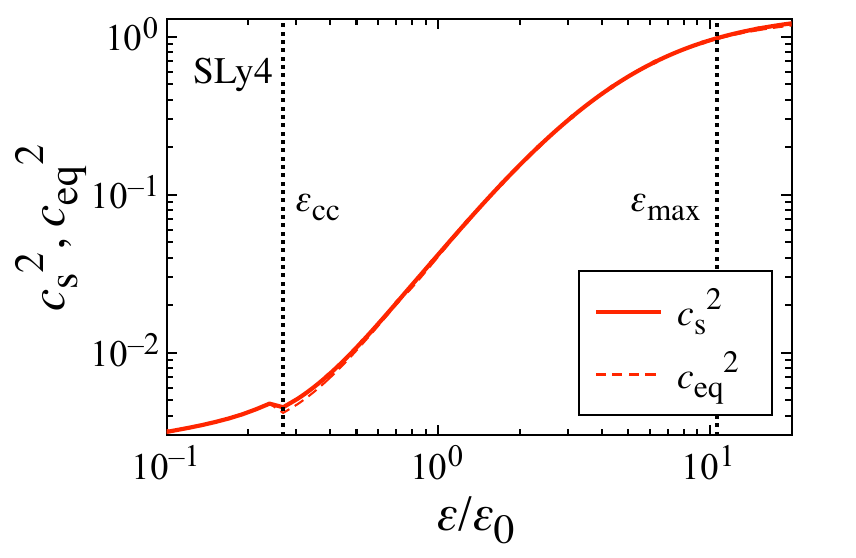} 
\end{center}
\caption{
The comparison between $c_{\rm s}^2$ and $c_{\rm eq}^2$ for SLy4. The horizontal axis is the energy density normalized by the saturation density, $\varepsilon_0=2.68\times 10^{14}$ g/cm$^3$. The dotted lines denote the transition density between the crust and core regions, $\varepsilon_{cc}$, and the density with which the stellar mass becomes maximum, $\varepsilon_{\rm max}$.
}
\label{fig:cs_SLy4}
\end{figure}

One of the important properties characterizing the EOS is sound velocity, $c_{\rm s}$, which is defined by
\begin{equation}
  c_{\rm s}^2 = \left(\frac{\partial p}{\partial \varepsilon}\right)_{s,Y_i}, \label{eq:cs}
\end{equation}
where $p$ and $\varepsilon$ are the pressure and energy density, respectively, while $s$ and $Y_i$ denote the entropy per baryon and the $i$-th composition fraction, respectively.  Here, $Y_i$ is given by
\begin{equation}
  Y_i = \frac{n_i}{n_{\rm b}}, \label{eq:Yi}
\end{equation}
where $n_i$ denotes the $i$-th composition number density, while $n_{\rm b}$ is the baryon number density, i.e., $n_{\rm b}=n_n+n_p$, where $n_n$ and $n_p$ denote the neutron and proton number densities, respectively. If the timescale of weak interactions is shorter than that of the stellar oscillations, the matter element that deviates from the equilibrium position settles into beta equilibrium in the new position, i.e., the matter element does not feel the $Y_i$ gradient. That is, one can consider that $c_{\rm s}$ is the same as the velocity, $c_{\rm eq}$, given by 
\begin{equation}
  c_{\rm eq}^2 \equiv \frac{dp/dr}{d\varepsilon/dr}, \label{eq:cs_eq}
\end{equation}
during the stellar oscillations. However, in general, the timescale of the weak interaction is longer than that of the stellar oscillations even for a cold neutron star~\cite{DH01}, and $c_{\rm s}$ should be considered to be different from $c_{\rm eq}$ because of the variation of $Y_i$ inside the star. In Fig.~\ref{fig:cs_SLy4}, as an example, we show the $c_{\rm s}^2$ and $c_{\rm eq}^2$ for SLy4 as a function of energy density normalized by the saturation density, $\varepsilon_0=2.68\times 10^{14}$ g/cm$^3$. From this figure, it is observed that the deviation of $c_{\rm s}^2$ from $c_{\rm eq}^2$ is significantly small compared to the absolute values of $c_{\rm s}^2$ or $c_{\rm eq}^2$, even though $c_{\rm s}^2$ is always slightly larger than $c_{\rm eq}^2$. 
We note that $c_{\rm s}^2$ is assumed to be identical to $c_{\rm eq}^2$ in the crust region, as in the previous studies. This is because the crust EOSs adopted in this study are available only for $\beta$-equilibrium matter, while the distinction between $c_{\rm s}$ and $c_{\rm eq}$ is explicitly evaluated only for the uniform core EOSs.

\begin{figure}[tbp]
\begin{center}
\includegraphics[scale=0.6]{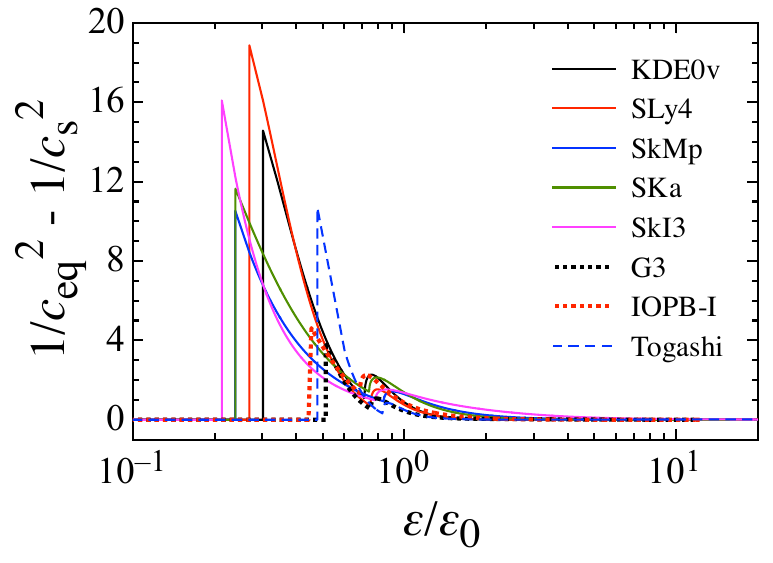} 
\end{center}
\caption{
The difference between $1/c_{\rm eq}^2$ and $1/c_{\rm s}^2$ for various EOSs is shown as a function of the energy density normalized by the saturation density. 
}
\label{fig:ceq_cs}
\end{figure}

In practice, the $g$-mode oscillations are associated with the Brunt-V\"{a}is\"{a}l\"{a} frequencies, $f_{\rm BV}$, which are criteria to judge the convectional (in)stability, i.e., convectionally stable if $f_{\rm BV}>0$. $f_{\rm BV}$ is locally determined by
\begin{equation}
  f_{\rm BV} = {\rm sgn}({\cal N}^2)\sqrt{|{\cal N}^2|}/(2\pi). \label{eq:BV}
\end{equation}
In this expression, ${\cal N}^2$ is given by
\begin{equation}
  {\cal N}^2 = -e^{2\Phi - 2\Lambda} \frac{\Phi'p'}{\varepsilon + p}\left(\frac{1}{c_{\rm eq}^2} - \frac{1}{c_{\rm s}^2}\right), \label{eq:N2}
\end{equation}
where the prime denotes the derivative with respect to $r$. In Fig.~\ref{fig:ceq_cs}, we show the difference of $1/c_{\rm eq}^2$ and $1/c_{\rm s}^2$ as a function of the energy density normalized by the saturation density, i.e., $\varepsilon_0=2.68\times 10^{14}$ g/cm$^3$ for various EOS. For any EOS, the difference between $1/c_{\rm eq}^2$ and $1/c_{\rm s}^2$ is zero in the crust region by definition and discontinuously increases at the transition density between the crust and core. Then, the difference between $1/c_{\rm eq}^2$ and $1/c_{\rm s}^2$ decreases as the density increases, but it increases again at the density where the muon appears.

\section{Eigenfrequencies}
\label{sec:Oscillations}

On the neutron star models constructed with the EOS mentioned in the previous section, we perform a linear analysis to determine the eigenfrequencies. In this study, we simply adopt the Cowling approximation, i.e., the metric remains fixed during the fluid oscillations. With this approximation, the perturbation equations are derived from a linearized energy-momentum conservation law. Then, by imposing the appropriate boundary conditions, the problem to solve becomes an eigenvalue problem. That is, one has to impose the regularity conditions at the center and the condition that the Lagrangian perturbation of pressure should vanish at the stellar surface. The concrete system of equations and the boundary conditions are the same as in Ref.~\cite{STM2001,ST2020}. We note that the perturbation equations include the term proportional to $1/c_{\rm eq}^2 - 1/c_{\rm s}^2$. In the case with $1/c_{\rm eq}^2 - 1/c_{\rm s}^2=0$, one can get only the $f$- and $p$-mode frequencies, while in the case with $1/c_{\rm eq}^2 - 1/c_{\rm s}^2\ne 0$ one can obtain the $g$-mode frequencies as well as the $f$- and $p$-mode frequencies by solving the same perturbation equations and boundary conditions. Namely, if the timescale of the weak interaction is shorter than that of the stellar oscillations, where ${\cal N}^2= 0$, the $g$-mode frequencies can not be excited. We also note that one should impose the junction conditions at the interface between crust and core regions, where we assume that $c_{\rm s}$ is discontinuous. That is, the Lagrangian displacement in the radial direction and the Lagrangian perturbation of pressure should be continuous at the transition density, which leads to the fact that the Lagrangian displacement in the tangential directions also becomes continuous at least in the Cowling approximation.

The accuracy of the Cowling approximation for the $g$-mode frequencies in cold neutron stars is less than $\sim 10 \%$, which becomes better for a lighter neutron star model~\cite{Zhao22}. Since the damping rate for the $g$-mode is generally too small, which leads to less gravitational radiation, one may determine the relatively accurate frequencies even without the metric perturbations.  

\begin{figure}[tbp]
\begin{center}
\includegraphics[scale=0.6]{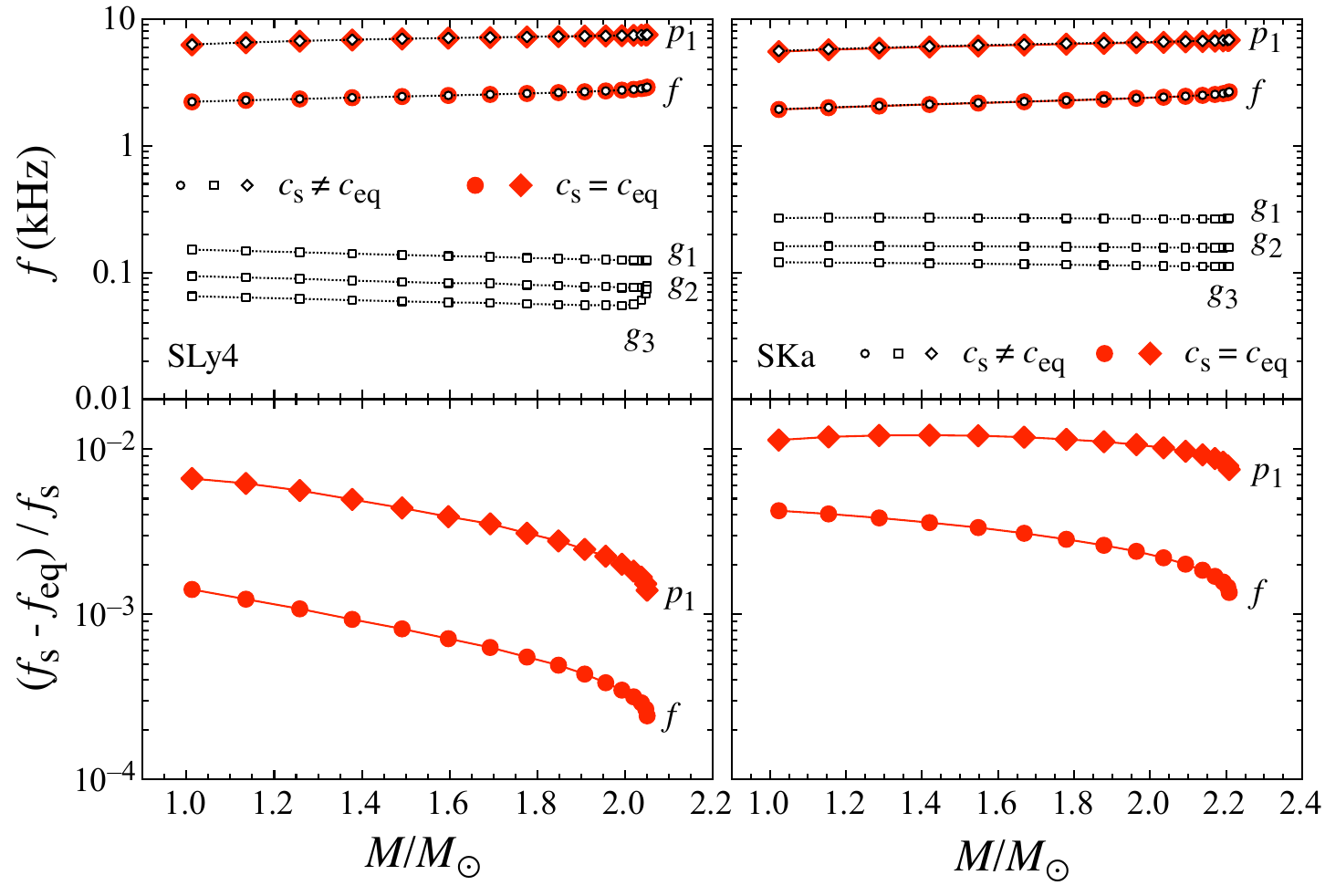} 
\end{center}
\caption{
The excited frequencies on the neutron star models constructed with SLy4 (top-left panel) and SKa (top-right panel) are shown as a function of the stellar mass. The open marks denote the frequencies, $f_{\rm s}$, determined with the sound velocity given by Eq.~(\ref{eq:cs}), i.e., $c_{\rm s}\ne c_{\rm eq}$, while the filled marks are those, $f_{\rm eq}$, determined with the assumption of $c_{\rm s}=c_{\rm eq}$. The circle, square, and diamond correspond to the $f$-, $g$-, and $p$-mode frequencies. The bottom panel shows the relative deviation between $f_{\rm s}$ and $f_{\rm eq}$.
}
\label{fig:fpg_SLy4}
\end{figure}

Before focusing on the $g$-mode frequencies, we show the eigenfrequencies excited in the neutron star models constructed with SLy4 and SKa as a function of the stellar mass in the top panel of Fig.~\ref{fig:fpg_SLy4}, where the open marks denote the frequencies, $f_{\rm s}$, calculated with the sound velocity defined by Eq.~(\ref{eq:cs}), i.e., $c_{\rm s}\ne c_{\rm eq}$, while the filled marks are those, $f_{\rm eq}$, with the assumption of $c_{\rm s}=c_{\rm eq}$. And, the circles, squares, and diamonds correspond to the $f$-, $g$-, and $p$-mode frequencies. Since, as mentioned before, the $g$-modes can not be excited in the neutron star models with the assumption of $c_{\rm s}=c_{\rm eq}$, the frequencies of $f_{\rm eq}$ are only $f$- and $p$-mode oscillations. From this figure, one can observe that the $g$-mode frequencies become larger as the EOS becomes stiffer, as shown in Ref.~\cite{Zhao22}. In the bottom panel of Fig.~\ref{fig:fpg_SLy4}, we also show the relative deviation between $f_{\rm s}$ and $f_{\rm eq}$. From this figure, one can also observe that the effect from the assumption that $c_{\rm s}$ is equivalent to $c_{\rm eq}$ seems to be almost negligible on the $f$- and $p$-modes. At the same time, this relative deviation becomes greater for overtones than for the fundamental oscillations. Therefore, when extremely high harmonics are taken into account, these deviations may become significant.

\begin{figure}[tbp]
\begin{center}
\includegraphics[scale=0.6]{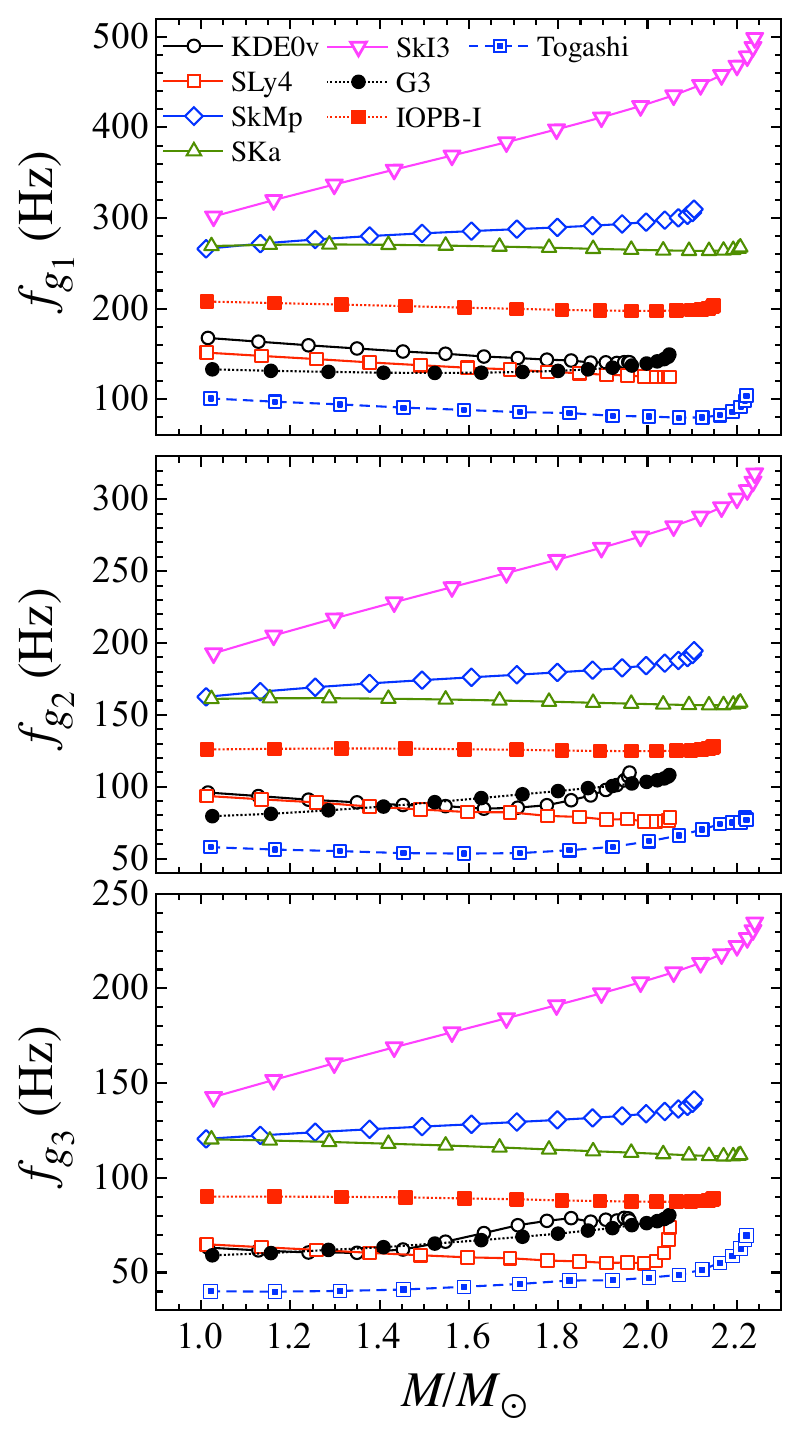} 
\end{center}
\caption{
The frequencies of the $g_i$-modes for $i=1,2,3$ excited in the neutron star models constructed with various EOS are shown as a function of the stellar mass, where the top, middle, and bottom panels respectively correspond to the $g_1$-, $g_2$-, and $g_3$-modes.}
\label{fig:g123_MR}
\end{figure}

Similarly, the $g_i$-mode frequencies for $i=1,2,3$ excited in the neutron star models constructed with various EOS listed in Table~\ref{tab:EOS}, are shown as a function of the stellar mass in Fig.~\ref{fig:g123_MR}, where the top, middle, and bottom panels correspond to the $g_1$-, $g_2$-, and $g_3$-modes. At a glance, the $g$-mode frequencies increase as the EOS becomes stiffer, e.g., compared to Togashi ($L=34.4$ MeV) with SkI3 ($L=101$ MeV). However, carefully observing the figure, the EOS dependence does not seem to be so simple, e.g., compared to SkMp ($L=70.3$ MeV) with SKa ($L=74.6$ MeV). In addition, the mass dependence of the frequencies strongly depends on the EOS model, i.e., the frequencies increase with the stellar mass for some EOS models, while those decrease for the other EOS models. Moreover, the behavior of the higher $g$-modes becomes more complicated than that of the $g_1$-modes. Thus, only using the relations between the $g$-mode frequencies and the stellar mass, it may be difficult to extract some information on the EOS properties, even if one would detect the corresponding frequencies in gravitational waves from an isolated neutron star.

\begin{figure}[tbp]
\begin{center}
\includegraphics[scale=0.6]{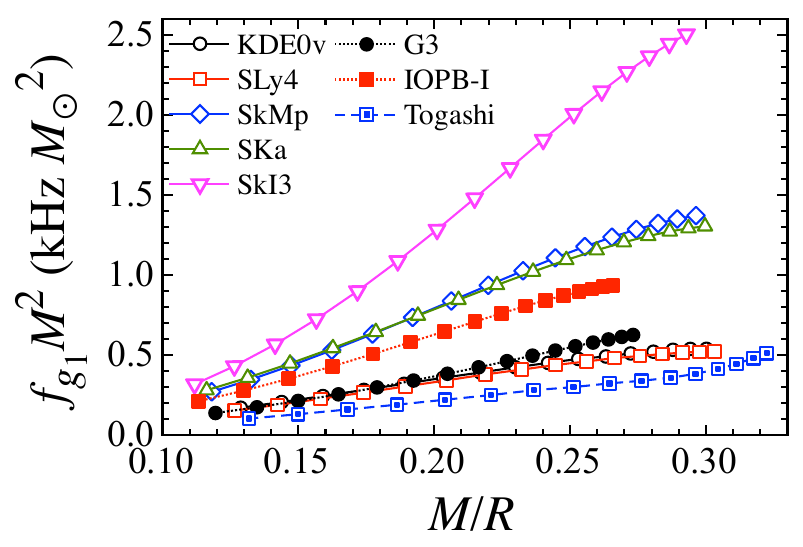} 
\end{center}
\caption{
The $g_1$-mode frequencies multiplied by the square of stellar mass, i.e., $f_{g_1}M^2$, are shown as a function of the stellar compactness, $M/R$, for various EOS listed in Table~\ref{tab:EOS}. 
}
\label{fig:fM2_MR}
\end{figure}

Nevertheless, we find that, if the $g_1$-mode frequencies multiplied by the square of the stellar mass, i.e., $f_{g_1}M^2$, are plotted as a function of the stellar compactness, $M/R$, as shown in Fig.~\ref{fig:fM2_MR}, the EOS dependence seems to become better than the relation between the $g_1$-mode frequencies and the stellar mass shown in the top panel of Fig.~\ref{fig:g123_MR}. In this case, $f_{g_1}M^2$ can be well expressed as a function of $M/R$ as
\begin{equation}
   f_{g_1}M^2\ ({\rm kHz}/M_\odot^2)
   = a_{10} 
   + a_{11}\left(\frac{M/R}{0.172}\right) 
   + a_{12}\left(\frac{M/R}{0.172}\right)^2 
   + a_{13}\left(\frac{M/R}{0.172}\right)^3, \label{eq:fit1}
\end{equation}
where the coefficients, $a_{1j}$ for $j=0-3$, depend on the EOS parameters. The normalization factor, 0.172, corresponds to the stellar compactness for the canonical neutron star model with $1.4M_\odot$ and 12 km. We note the coefficients of $a_{1j}$ for $j=0-3$ are in the unit of kHz/$M_\odot^2$. Furthermore, we find that these coefficients, $a_{1j}$, can be expressed as a function of the dimensionless parameter given by $\eta_0\equiv L/K_0$:
\begin{gather}
   a_{10} = -8.2274 + 22.9039\bar{\eta}_0 -20.4412\bar{\eta}_0^2 + 6.1178\bar{\eta}_0^3, \label{eq:fit2a} \\
   a_{11} = 21.4329 -59.3407\bar{\eta}_0 + 52.8381\bar{\eta}_0^2 -16.0556\bar{\eta}_0^3, \label{eq:fit2b} \\
   a_{12} = -17.1318 + 46.6050\bar{\eta}_0 -40.0223\bar{\eta}_0^2 + 12.2336\bar{\eta}_0^3, \label{eq:fit2c} \\
   a_{13} = 4.4126 -11.6211\bar{\eta}_0 + 9.5167\bar{\eta}_0^2 -2.8029\bar{\eta}_0^3, \label{eq:fit2d}   
\end{gather}
where $\bar{\eta}_0$ is $\eta_0$ normalized by $0.25$, i.e., $\bar{\eta}_0\equiv\eta_0/0.25$. The normalization factor, 0.25, is the value of $\eta_0$ using $K_0=240$ and $L=60$ MeV. 

\begin{figure}[tbp]
\begin{center}
\includegraphics[scale=0.6]{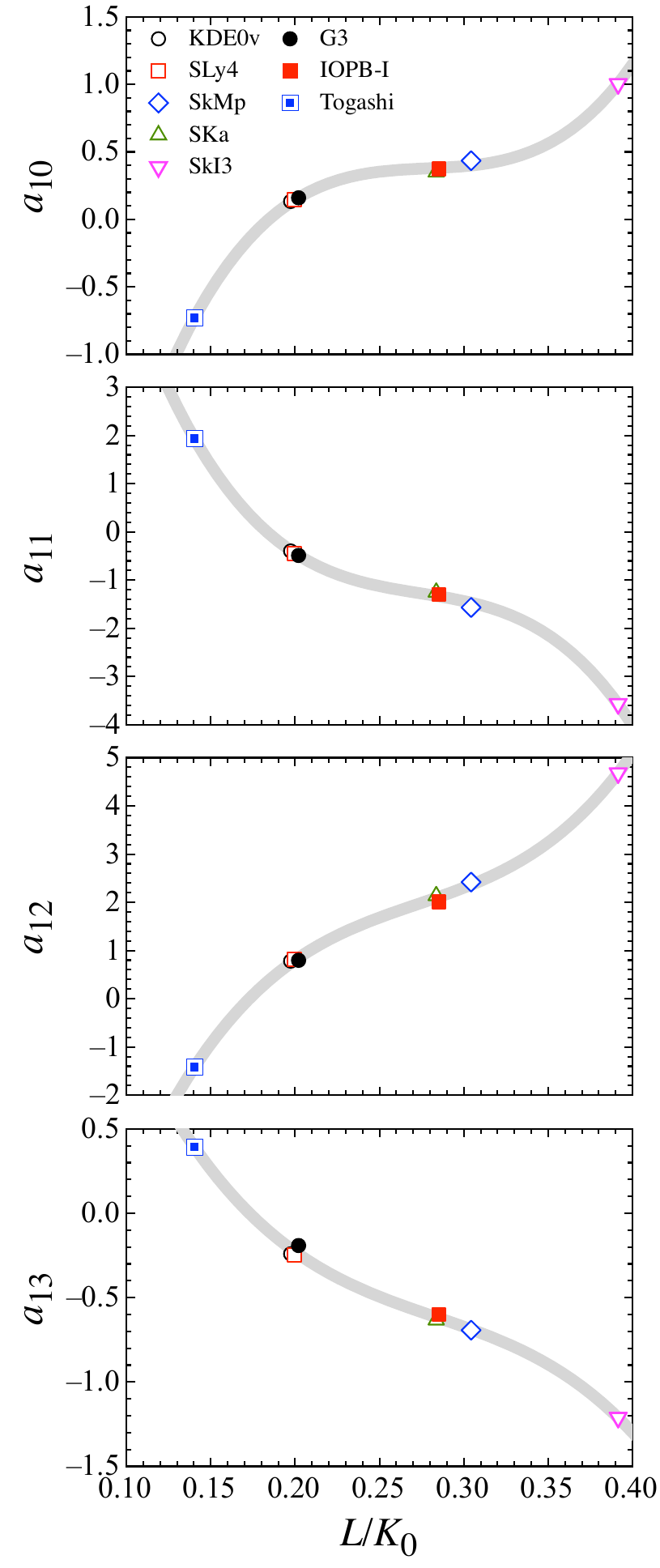} 
\end{center}
\caption{
For $g_1$-modes, the coefficients, $a_{1j}$, in Eq.~(\ref{eq:fit1}) are shown as a function of $L/K_0(\equiv \eta_0)$ for various EOS considered in this study. The thick solid lines denote the fitting lines given by Eq.~(\ref{eq:fit2a})-(\ref{eq:fit2d}). 
}
\label{fig:aix}
\end{figure}

\begin{figure}[tbp]
\begin{center}
\includegraphics[scale=0.6]{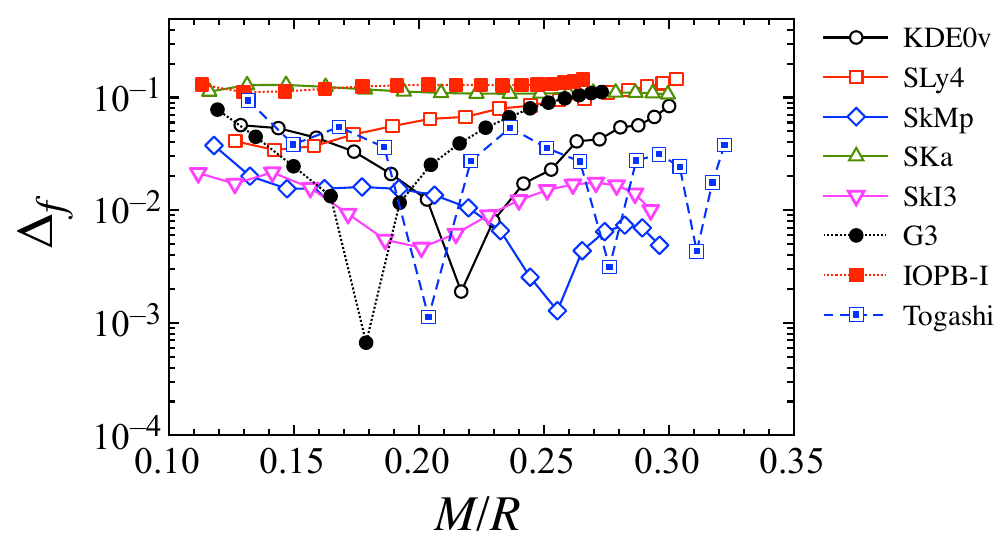} 
\end{center}
\caption{
Relative deviation of the frequencies estimated with the fitting formula from the frequencies determined from the eigenvalue problem, defined by Eq.~(\ref{eq:Delta}), is shown as a function of the stellar compactness.
}
\label{fig:deltaf_MR}
\end{figure}

In practice, to see how well our fitting formula for the $g_1$-mode frequencies given by Eqs.~(\ref{eq:fit1}) - (\ref{eq:fit2d}) works, in Fig.~\ref{fig:deltaf_MR}, we show the relative deviation defined by
\begin{equation}
  \Delta_f \equiv \frac{|f_{g_1}^{(\rm eigen)} - f_{g_1}^{(\rm fit)}|}{f_{g_1}^{(\rm eigen)}}, \label{eq:Delta}
\end{equation}
where $f_{g_1}^{(\rm eigen)}$ and $f_{g_1}^{(\rm fit)}$ are, respectively, the $g_1$-mode frequencies obtained by solving the eigenvalue problem and those estimated using the fitting formulae. From this figure, one can observe that our fitting formula predicts the $g_1$-mode frequencies with $\sim 10\%$ accuracy. 
For all 128 stellar models (16 neutron-star models for each of the 8 EOS), we also evaluate the residuals between the $g_1$-mode frequencies obtained from the eigenvalue problem and those predicted with the fitting formula. We find that the mean residual is essentially zero, and its standard deviation is 0.0156 kHz, with $71.9\%$ and $96.9\%$ of the models lying within $1\sigma$ and $2\sigma$, respectively. This result is consistent with a normal distribution ($68.3\%$ and $95.4\%$), indicating that our fitting formula reproduces the numerical results without significant systematic bias.

We have to emphasize that our fitting formula for the $g_1$-mode frequencies as a function of $M/R$ and $\eta_0$ was found through trial and error, 
assuming that $f_{g_1}M^k$ as a function of $M/R$ and $\eta_0=(K_0^iL^j)^{1/(i+j)}$ with integers of $i$, $j$, and $k$,
so it may not be the only one. Namely, another functional form may exist for expressing the $g_1$-mode frequencies, and it may be better than our fitting formula.
Since the $g_1$-mode frequencies are well expressed as our empirical formula, the combination of $f_{g_1}M^2$ and/or $\eta_0$ may have some physical meaning. But, unfortunately, we could not find such a physical background. 
In addition, we consider similarly deriving the fitting formula for the $g_i$-mode frequencies for $i\ge 2$, but we could not find them, where the frequencies can not be well fitted as a cubic function of $M/R$ as in Eq.~(\ref{eq:fit1}). This may come from the fact that the nodal number inside the star contained in the eigenfunction of the $g_i$-mode is $i$. As the nodal number increases, the eigenfunction is easily affected by the interior properties, such as the radial profile of the sound velocity. This is the same as in the $p_i$-mode frequencies.

\section{Dependence of the $g$-mode frequencies on the sound velocity}
\label{sec:Dep_cs}

\begin{figure}[tbp]
\begin{center}
\includegraphics[scale=0.6]{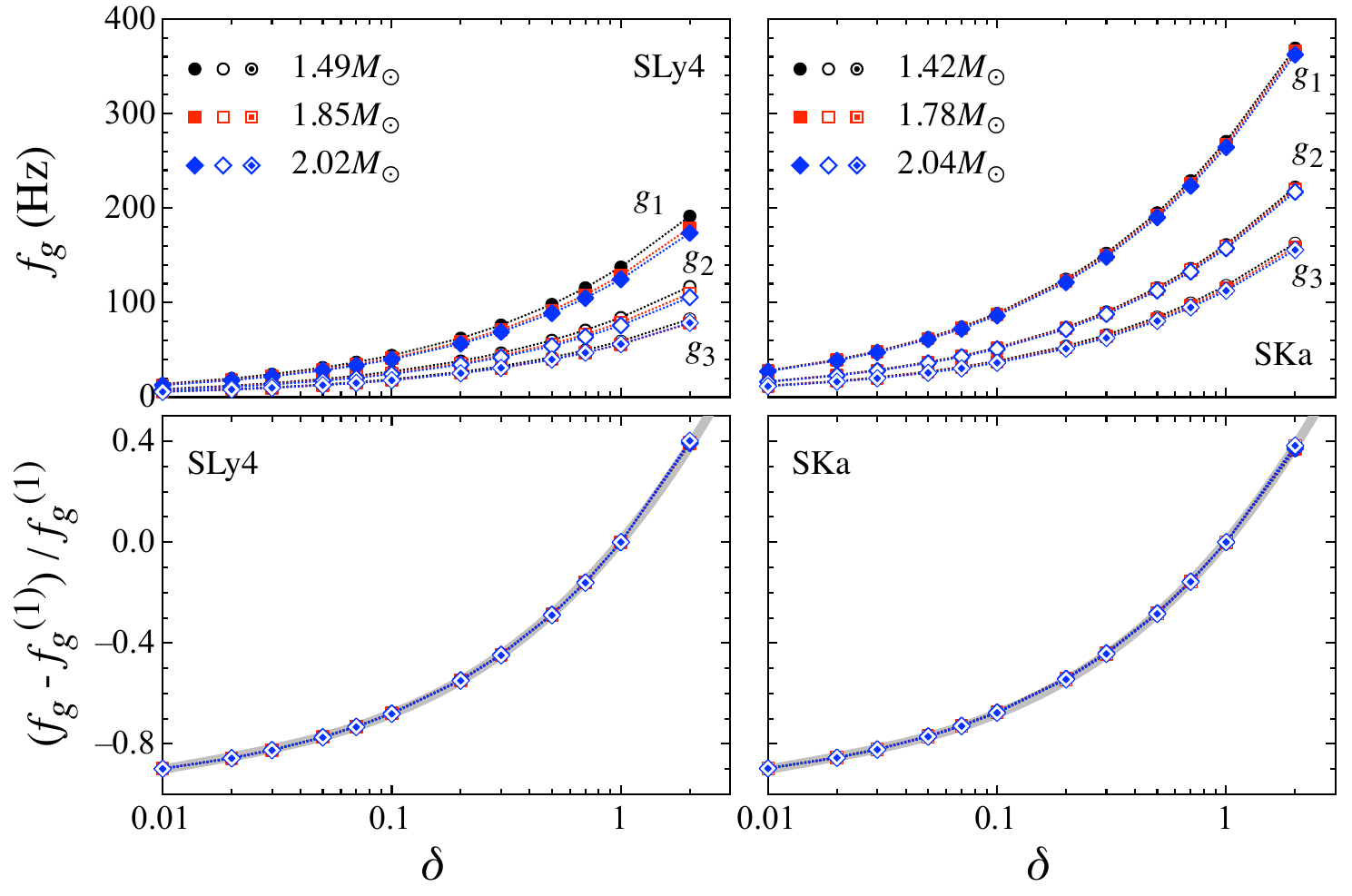} 
\end{center}
\caption{
In the top panels, we show the frequencies of the $g_1$-mode (filled marks), $g_2$-mode (open marks), and $g_3$-mode (double marks) as a function of $\delta$ for the neutron star models constructed with SLy4 (left panel) and SKa (right panel), where the circles, squares, and diamonds denote the results for the neutron star models with different mass. The bottom panels are the relative deviation of the $g$-mode frequencies, $f_g$, with various values of $\delta$ from that with $\delta=1$, $f_g^{(1)}$. In addition, the thick solid line in the bottom panel is the fitting formula given by Eq.~(\ref{eq:fit_devi}).
}
\label{fig:dep_delta}
\end{figure}

It is known that the $g$-mode frequencies disappear in the limit of $c_{\rm s}=c_{\rm eq}$, as mentioned before. However, it is unclear how the $g$-mode frequencies disappear when $c_{\rm s}$ approaches $c_{\rm eq}$. To see this behavior, we introduce the dimensionless variable, $\delta$, with which the sound velocity is artificially modified as 
\begin{equation}
  \bar{c}_{\rm s}^2 = c_{\rm eq}^2 + \delta \left(c_{\rm s}^2 - c_{\rm eq}^2\right), \label{eq:d_cs2}
\end{equation}
with $c_{\rm s}$ and $c_{\rm eq}$ being the velocities defined with Eqs.~(\ref{eq:cs}) and (\ref{eq:cs_eq}). That is, the case with $\delta=1$ is $\bar{c}_{\rm s}=c_{\rm s}$, while the case with $\delta=0$ is $\bar{c}_{\rm s}=c_{\rm eq}$. Then, we examine how the frequencies change when $c_{\rm s}$ and $c_{\rm eq}$ are chosen as the same as in the given EOS, while $\bar{c}_{\rm s}^2$ with various values of $\delta$ is adopted instead of $c_{\rm s}^2$ in the perturbation equations. In the top panels of Fig.~\ref{fig:dep_delta}, we show how the $g_i$-mode frequencies with $i=1,2,3$ depend on the value of $\delta$, where the left and right panels correspond to the results for the stellar models constructed with SLy4 and SKa. In this figure, the filled, open, and double marks respectively denote the $g_1$-, $g_2$-, and $g_3$-modes, while the circles, squares, and diamonds correspond to the results for the neutron star with different masses. From this figure, one can observe that the $g$-mode frequencies strongly depend on the value of $\delta$, i.e., the value of $c_{\rm s}$. In the bottom panels, we also show the relative deviation of the $g$-mode frequencies with various values of $\delta$ from that with $\delta=1$ (appropriate value of $c_s$). In both bottom panels, the thick solid line is the fitting line given by
\begin{equation}
  \left(f_g-f_g^{(1)}\right)/f_g^{(1)} = -0.6806 + 0.3620 \chi + 0.2318\chi^2 +  0.090332\chi^3, \label{eq:fit_devi}
\end{equation}
where $\chi\equiv \log_{10}\left(\delta/0.1\right)$. From this figure, we surprisingly find that the relative deviation hardly depends on the neutron star mass, EOS, and also the nodal number in the eigenfunction of $g_i$-modes. We find that the $g$-mode frequencies are reduced by $\sim 23\%$ with $\delta = 0.5$, while they increase by $\sim 39\%$ with $\delta = 2$.

\section{Conclusion}
\label{sec:Conclusion}

When the sound velocity, $c_{\rm s}$, is properly accounted for, the square of the sound velocity deviates from the pressure gradient, $c_{\rm eq}=dp/d\varepsilon$, along the $\beta$ equilibrium. Due to this deviation, the $g$-mode oscillations are also excited even in a zero-temperature neutron star. On the other hand, the $f$- and $p_1$-mode frequencies are almost unchanged even if one takes into account the difference between $c_{\rm s}$ and $c_{\rm eq}$. Although the frequency of the $g$-mode tends to increase as the stiffness of the EOS increases, the behavior is not quite that simple. Nevertheless, we find a fitting formula expressing the 1st $g$-mode ($g_1$-mode) frequencies as a function of the stellar mass, compactness, $M/R$, and the combination of the nuclear saturation parameter, $\eta_0\equiv L/K_0$. Using our fitting formula, one can evaluate the $g_1$-mode frequencies with $\sim 10\%$ accuracy. However, since this rule of thumb was derived through trial and error, there may be a better one out there. We also tried to find the fitting formula for the $g_i$-mode frequencies for $i\ge 2$, but unfortunately, that attempt failed. If one could identify a specific frequency in the gravitational waves from a neutron star with the $g_1$-mode frequency, one may be able to estimate the value of $\eta_0$ using our fitting formula. Or, 
although the additional compositions, such as hyperons and/or quarks, may appear in realistic neutron star models, if we observe the deviation of the $g_1$-mode frequencies from our fitting formula, we may infer the presence of such an additional composition inside the star. Furthermore, we examined the behavior of the $g$-mode frequencies, when the deviation between $c_{\rm s}$ and $c_{\rm eq}$ artificially varies according to Eq.~(\ref{eq:d_cs2}), introducing a new parameter $\delta$ characterizing the deviation between $c_{\rm s}$ and $c_{\rm eq}$, which shows that the $g$-mode frequencies are quite sensitive to the determination of the sound velocity. 

In this study, since we adopted the Cowling approximation to derive the empirical formulas, the coefficients in such formulas have to be modified after the determination of the frequencies with the metric perturbations. Nevertheless, we have to emphasize that since the qualitative behavior of the frequencies determined with the Cowling approximation is almost the same as that with the metric perturbations, owing to our finding in this study, one can derive the empirical relation without finding a suitable combination of the stellar bulk properties and nuclear saturation parameters.
In addition, in this study, we have determined the oscillation frequencies with a linear perturbation analysis, with which we cannot determine the amplitude of each oscillation mode. That is, we cannot discuss the gravitational wave energy carried by each mode, directly associated with its detectability. To determine the gravitational radiation energy, we have to perform the nonlinear analysis, even though the radiation energy still depends on the initial conditions. On the other hand, considering that the $g$-mode frequencies are in a relatively lower frequency range, i.e., $\lesssim 600$ Hz, this frequency band is much better than that for the fundamental oscillation. So, the detection of gravitational waves with $g$-mode oscillations is still challenging, but may be possible if the radiative energy is sufficient, with the next-generation detectors, such as the Einstein Telescope and Cosmic Explorer. In the end, we also mention the effect of the rotation and magnetic field on the $g$-mode frequencies. Since the Alfv\'{e}n frequencies excited with magnetic field, $\omega_A$, is estimated that $\omega_A\sim v_A/R$ with the Alfv\'{e}n velocity given by $v_A=B/\sqrt{4\pi (\varepsilon+p)}$ and the stellar radius $R$, $\omega_A$ becomes $\sim 100$ Hz with $B\gtrsim 10^{16}$ G, e.g., Ref. \cite{Sotani09}. Thus, the magnetic effect on the $g$-mode frequencies is likely to be negligible even for a magnetar with $B=10^{15}$ G. Meanwhile, the rotational effect may not be that simple. When a stellar rotation is slow enough that $2\Omega \ll \omega_g$ holds with angular velocity $\Omega$, the rotational effect on the $g$-mode is almost negligible or can modify as $\omega_g=\omega_g^{(0)}+m\alpha \Omega$, where $\omega_g$, $\omega_g^{(0)}$, $m$, and $\alpha$ are the $g$-mode frequencies including the rotational effect, the $g$-mode frequencies without stellar rotation, azimuthal quantum number, and dimensionless rotation correction coefficient, e.g., Ref.~\cite{Lee96}. If the stellar rotation becomes so fast that the Coriolis force becomes comparable to the buoyancy force, one has to consider the rotational correction of the $g$-mode, i.e., an inertial-gravity mode. If the star's rotation were even faster and the buoyancy force could be neglected, the $g$-modes eventually change to the inertial modes~\cite{Passamonti09,Erich11}. Anyway, the empirical relationship identified in this study could likely be extended to account for rotational and magnetic effects.

\acknowledgments

We are grateful to Ankit KUMAR for preparing the EOS data of G3 and IOPB-I for this study. This work is supported in part by Japan Society for the Promotion of Science (JSPS) KAKENHI Grant Numbers 
JP23K20848,  
JP24KF0090, 
JP21K13924, and JP26K07085 
and by the Mitsubishi Foundation through grant No. 202510029.



\end{document}